 \documentclass{aa}  
\usepackage{graphicx}
\usepackage{txfonts}
\begin{document}

   \title{Detection of X-ray flares from AX~J1714.1$-$3912,
   the unidentified source near RX~J1713.7$-$3946}

%   \subtitle{Identification as a Supergiant Fast X-ray Transient}

   \author{Marco Miceli
          \inst{1,2}
          \and
          Aya Bamba\inst{3,4}}

   \institute{Dipartimento di Fisica \& Chimica, Università di Palermo, Piazza del Parlamento 1, 90134 Palermo, Italy\\
              \email{miceli@astropa.unipa.it}
         \and
             INAF-Osservatorio Astronomico di Palermo, Piazza del Parlamento 1,
             90134, Palermo, Italy
         \and
             Department of Physics, Graduate School of Science,
             The University of Tokyo, 7-3-1 Hongo, Bunkyo-ku, Tokyo, 113-0033,
             Japan
         \and
             Research Center for the Early Universe, School of Science,
             The University of Tokyo, 7-3-1 Hongo, Bunkyo-ku, Tokyo, 113-0033,
             Japan
             }

  \date{Received September 15, 1996; accepted March 16, 1997}

\titlerunning{X-ray flares from AX~J1714.1$-$3912}
\authorrunning{Miceli and Bamba}

\abstract{}{}{}{}{} 
% 5 {} token are mandatory
 \abstract
   {Molecular clouds are predicted to emit nonthermal X-rays when they are close to particle-accelerating supernova remnants (SNRs), and the hard X-ray source AX~J1714.1$-$3912, near the SNR RX~J1713.7$-$3946, has long been considered a candidate for diffuse nonthermal emission associated with cosmic rays diffusing from the remnant to a closeby molecular cloud.}
   {We aim at ascertaining the nature of this source by analyzing two dedicated X-ray observations performed with {\it Suzaku} and {\it Chandra}.}
   {We extracted images from the data in various energy bands, spectra, and light curves and studied the long-term evolution of the X-ray emission on the basis of the $\sim4.5$ yr time separation between the two observations.}
   {We found that there is no diffuse emission associated with AX~J1714.1$-$3912, which is instead the point-like source CXOU J171343.9$-$391205. We discovered rapid time variability (timescale $\sim 10^3$ s), together with a high intrinsic absorption and a hard nonthermal spectrum (power law with photon index $\Gamma \sim 1.4$). We also found that the X-ray flux of the source drops down by 1--2 orders of magnitude on a timescale of a few years.}
   {Our results suggest a possible association between AX~J1714.1$-$3912 and a previously unknown supergiant fast X-ray transient, although further follow-up observations are necessary to prove this association definitively.}


   \maketitle
%
%-------------------------------------------------------------------

\section{Introduction}

Supernova remnant (SNR) shocks have been proved to be efficient sites of particle acceleration and can account for the observed spectrum of cosmic rays up to the knee at $\sim3\times10^{15}$ eV \citep{be87,bv07}, although conclusive evidence for the presence of PeV particles in SNRs is still lacking. 

Bright X-ray synchrotron emission from electrons accelerated up to TeV energies has been observed in the expanding shock of many young SNRs (see the reviews by \citealt{rey08} and \citealt{vin12}). Additional X-ray nonthermal emission is expected when a SNR interacts with a molecular cloud. \citet{bce00} showed that, in this case, the radiative shock transmitted through the cloud can accelerate the thermal pool of electrons from the cloud in the ionized shock precursor, thus producing bright synchrotron and bremsstrahlung emission. \citet{gac09} showed that the cloud (even if it is not interacting with a nearby remnant) can be an intense source of nonthermal X-rays (via bremsstrahlung and synchrotron radiations) and $\gamma-$ray emission due to high-energy particles that leave the acceleration site and diffusively reach the cloud. We then expect to observe bright nonthermal X-rays from molecular clouds in the proximity of an accelerating SNR. Nevertheless, this X-ray emission has not been confirmed yet.

The galactic SNR RX~J1713.7$-$3946 is an ideal target for this kind of study since it is a very efficient particle accelerator: this SNR is dominated by synchrotron emission in X-rays and is also a very bright {\it H.E.S.S.} source in gamma-rays (see \citealt{abd09} and references therein). The remnant is surrounded by dense molecular clouds and its distance is $\sim1$ kpc (\citealt{cdb04,mtt05,fss12}).

AX~J1714.1$-$3912 was discovered during the {\it ASCA} mapping of the SNR RX~J1713.7$-$3946, beyond the northern border of the shell \citep{uchiyama2002}. Its very flat spectrum with photon index $\Gamma\sim$1 is unusual for nonthermal X-ray emission from SNRs, which typically have photon indices of 2--3 \citep{matsumoto2007}. The source position coincides with a molecular cloud with high CO($J$=2--1)/CO($J$=1--0), and the {\it EGRET} unidentified source 3EG~J1714$-$3857 \citep{butt2001}.
With these observational facts, \citet{uchiyama2002} concluded that the flat X-ray emission originates from bremsstrahlung emission from subrelativistic protons or mildly relativistic electrons accelerated at the shock front of RX~J1713.7$-$3946 and propagating through the nearby, dense molecular cloud.
This would provide an important clue of the particle injection to the diffusive shock acceleration mechanism in SNR shocks.

On the other hand, the source is too compact to resolve its morphology with {\it ASCA}. Thus, we could not distinguish whether AX~J1714.1$-$3912 is truly diffuse or is a compact source. Compact X-ray sources such as X-ray binaries often show very hard X-ray emission and rapid time variability. Since AX~J1714.1$-$3912 is located close to the Galactic center, it is natural to expect contamination of unrelated point-like sources such as X-ray binaries. Finding new point-like sources is important by itself to accumulate samples, especially for rare populations.
We thus conducted {\it Suzaku} and {\it Chandra} follow-up observations to check the spatial extent and time variability of AX~J1714.1$-$3912.

In this paper, we report our results, which show the detection of a significant time variability and the association of AX~J1714.1$-$3912 with a point-like source.
We describe the observations and data reduction in \S\ref{sec:obs}, and results in \S\ref{sec:results}. \S\ref{sec:discussion} discusses on the origin of AX~J1714.1$-$3912.

\begin{figure*}[!htb]
\centering
\includegraphics[width=\columnwidth]{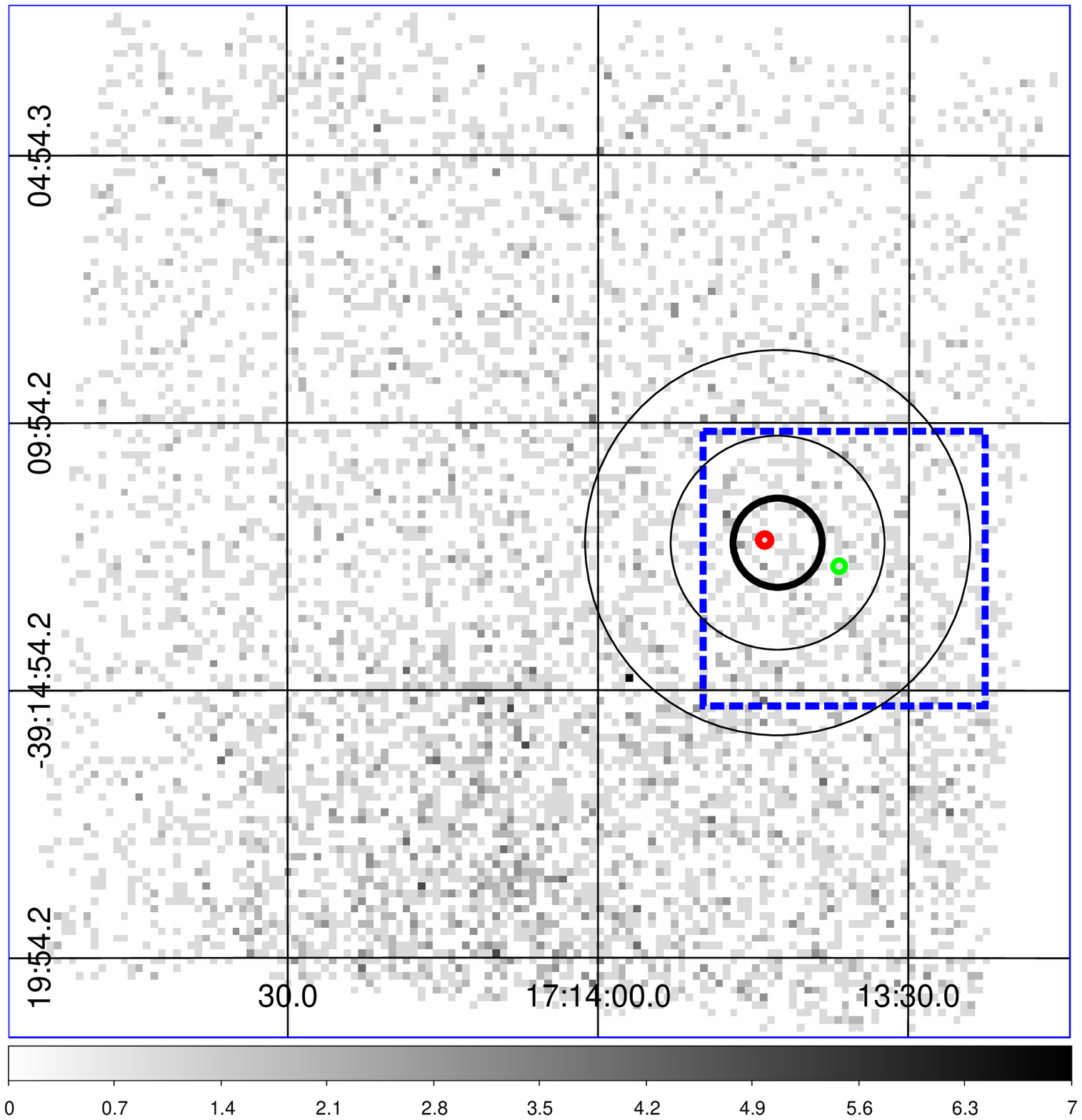}
\includegraphics[width=\columnwidth]{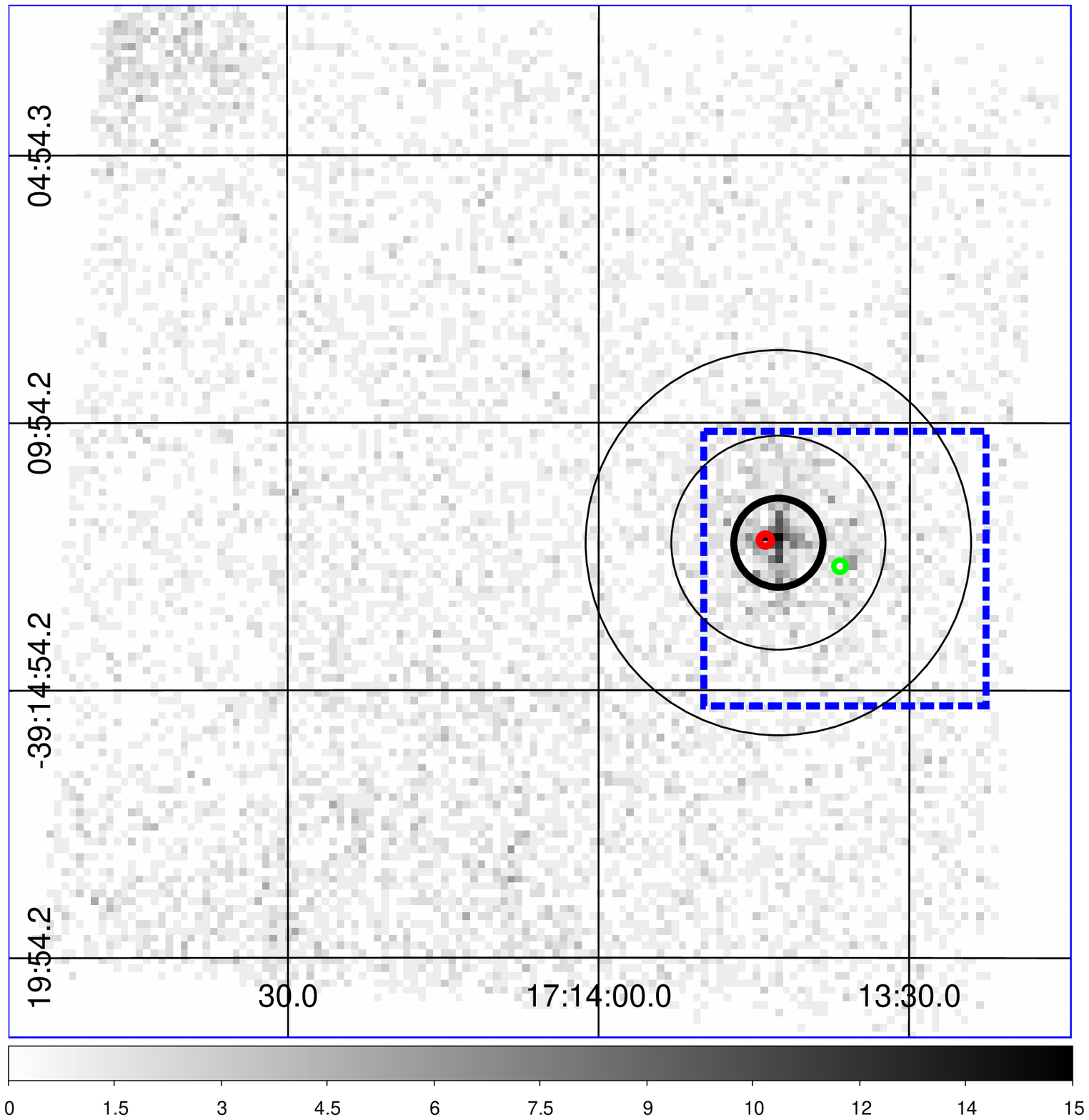}
\caption{{\it Suzaku} XIS3 0.5--2~keV (left) and 2--10~keV (right) image of the AX~J1714.1$-$3912 region in J2000 coordinates. The grayscale is in the linear scale in the unit of counts~pixel$^{-1}$. Thick and thin circles indicate source and background regions for timing and spectral analysis. The blue dashed box indicates the field of view of Fig. \ref{fig:chandraImage}. Red and green circles indicate the position of CXOU J171343.9$-$391205 and CXOU J171236.7$-$391235, respectively.}
\label{fig:SuzakuImage}
\end{figure*}

\section{Observations and data reduction}
\label{sec:obs}

%\subsection{{\it Suzaku}}
We analyzed {\it Suzaku} observation 505076010 (PI M. Miceli, $32.6$ ks of exposure time) performed on February 16, 2011, with pointing coordinates $\alpha_{J2000}=17^h14^m06.0^s$, $\delta_{J2000}=-39^{\circ}11'49.9"$. Data reduction and analysis were made with HEADAS software version 6.13. The data were reprocessed with the calibration database version XIS-20101108, XRT-20110630, and HXD-20101202. We performed the energy scale reprocessing by adopting the XISPI tool and followed the standard criteria for the data screening. 
We also analyzed {\it Chandra} ACIS-I observation 16767 (PI M. Miceli, $10.7$ ks of exposure time) performed on October 7, 2015 with pointing coordinates $\alpha_{J2000}=17^h13^m42.8^s$, $\delta_{J2000}=-39^{\circ}12'15.4"$. Data were reprocessed with CIAO 4.9 and CALDB 4.7.3. Images and spectra were extracted with the SPECEXTRACT and FLUXIMAGE scripts, respectively.

Spectral analysis was performed using XSPEC v12.9 \citep{arn96}. All the reported errors are at 90\% confidence level.

\section{Results}
\label{sec:results}

\subsection{{\it Suzaku}}
\label{Suzaku}

Figure~\ref{fig:SuzakuImage} shows the {\it Suzaku} 0.5--2~keV (left) and 2--10~keV (right) images of the AX~J1714.1$-$3912 region. We only used XIS3 data, since the image by XIS0 was partly affected by the anomaly\footnote{https://heasarc.gsfc.nasa.gov/docs/suzaku/news/xis0.html} and XIS1 has higher background level.
We can see a clear compact source on the position of AX~J1714.1$-$3912 only in the hard X-ray band image, implying that the source emission is deeply absorbed.
The peak position is at $\alpha_{J2000}=17^h13^m42.69^s$, $\delta_{J2000}=39^{\circ}12'8.7"$ in J2000 coordinates. We can also see a faint compact source on the right-hand side of our target with peak position $\alpha_{J2000}=17^h13^m35.78^s$, $\delta_{J2000}=39^{\circ}12'32.6"$, and the diffuse emission of the SNR RX~J1713$-$3946 on the south in both images. Sect. \ref{chandra} shows a more detailed description of the position and morphology of these two sources, as revealed by {\it Chandra}.

\begin{figure}[!htb]
\centering
\includegraphics[width=\columnwidth]{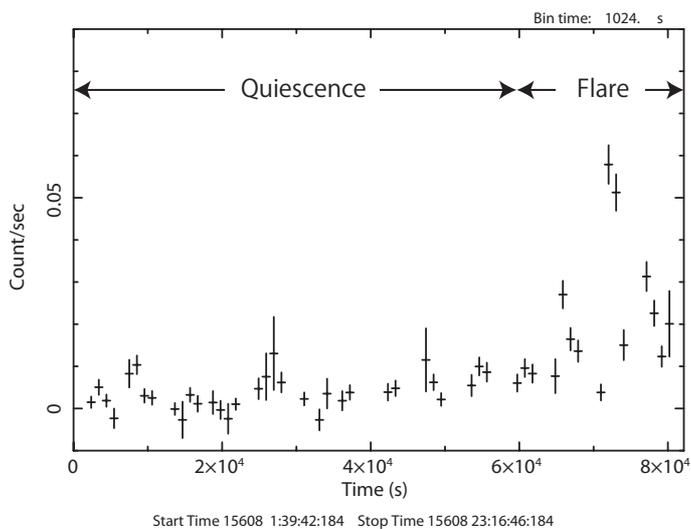}
\caption{Background-subtracted {\it Suzaku} XIS light curve of AX~J1714.1$-$3912 in the 2--10~keV band.}
\label{fig:SuzakuLC}
\end{figure}

Spectra and the light curve of AX~J1714.1$-$3912 are extracted from a circular region with a radius of 50~arcsec (thick black circle in Fig. \ref{fig:SuzakuImage}), which does not include the peak of the emission of the bright nearby source, whereas background photons are extracted from the annulus region with inner and outer radii of 2 and 3.6~arcmin, respectively (thin circles in Fig.~\ref{fig:SuzakuImage}).
Figure~\ref{fig:SuzakuLC} shows the background-subtracted light curve of AX~J1714.1$-$3912 in the 2--10~keV energy band with a timing bin of 1024~sec (all the three XIS data were averaged).
We can see clear flares at the end of the observation. By fitting the light curve with a constant, we obtain  $\chi^2$/d.o.f. of 1241.7/47, and the Kolmogorov-Smirnov test shows that the probability of constancy is less than $1\times 10^{-38}$. These tests confirm the significant time variability of the source.

\begin{figure*}[!htb]
\centering
\includegraphics[width=0.33\textwidth]{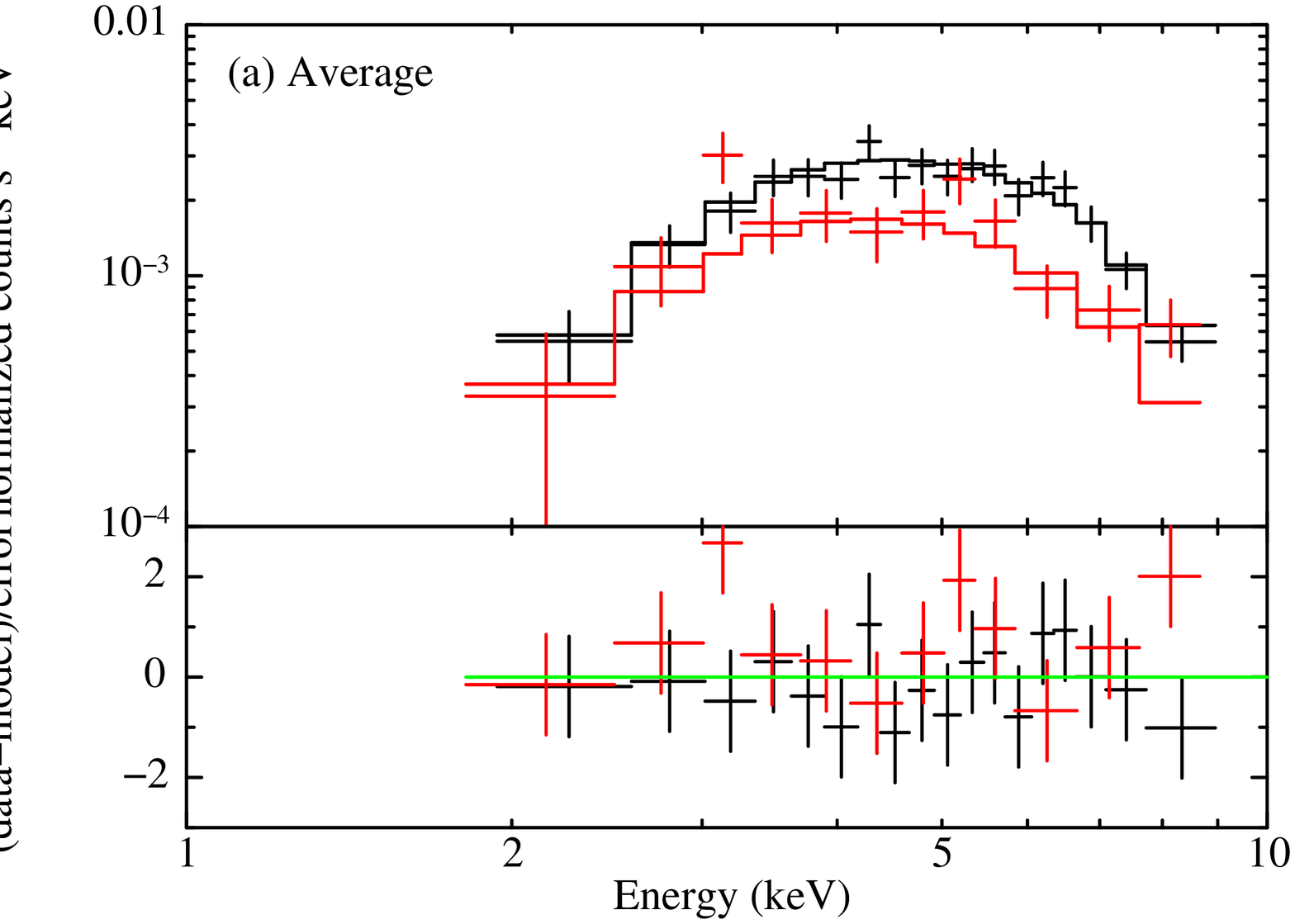}
\includegraphics[width=0.33\textwidth]{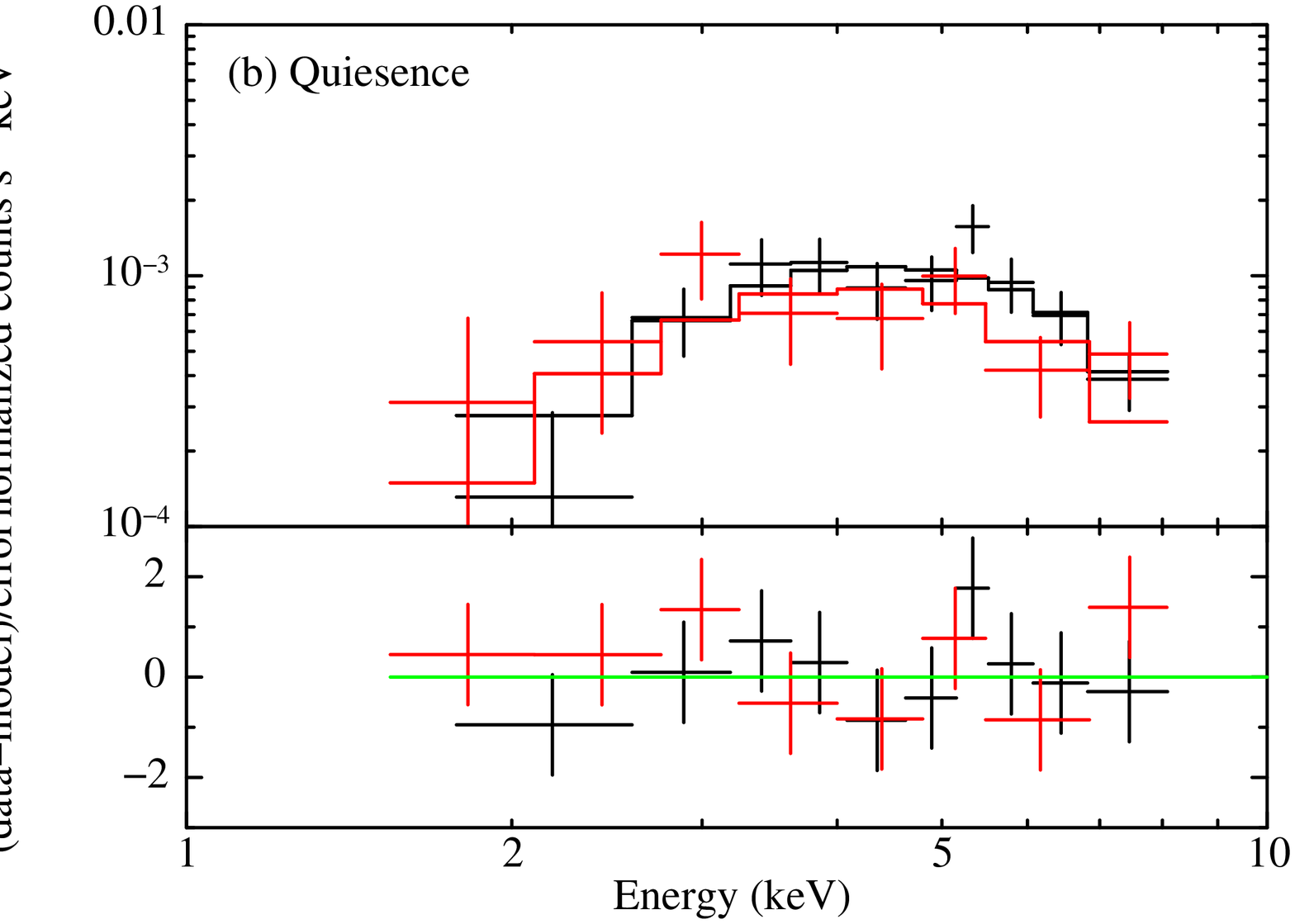}
\includegraphics[width=0.33\textwidth]{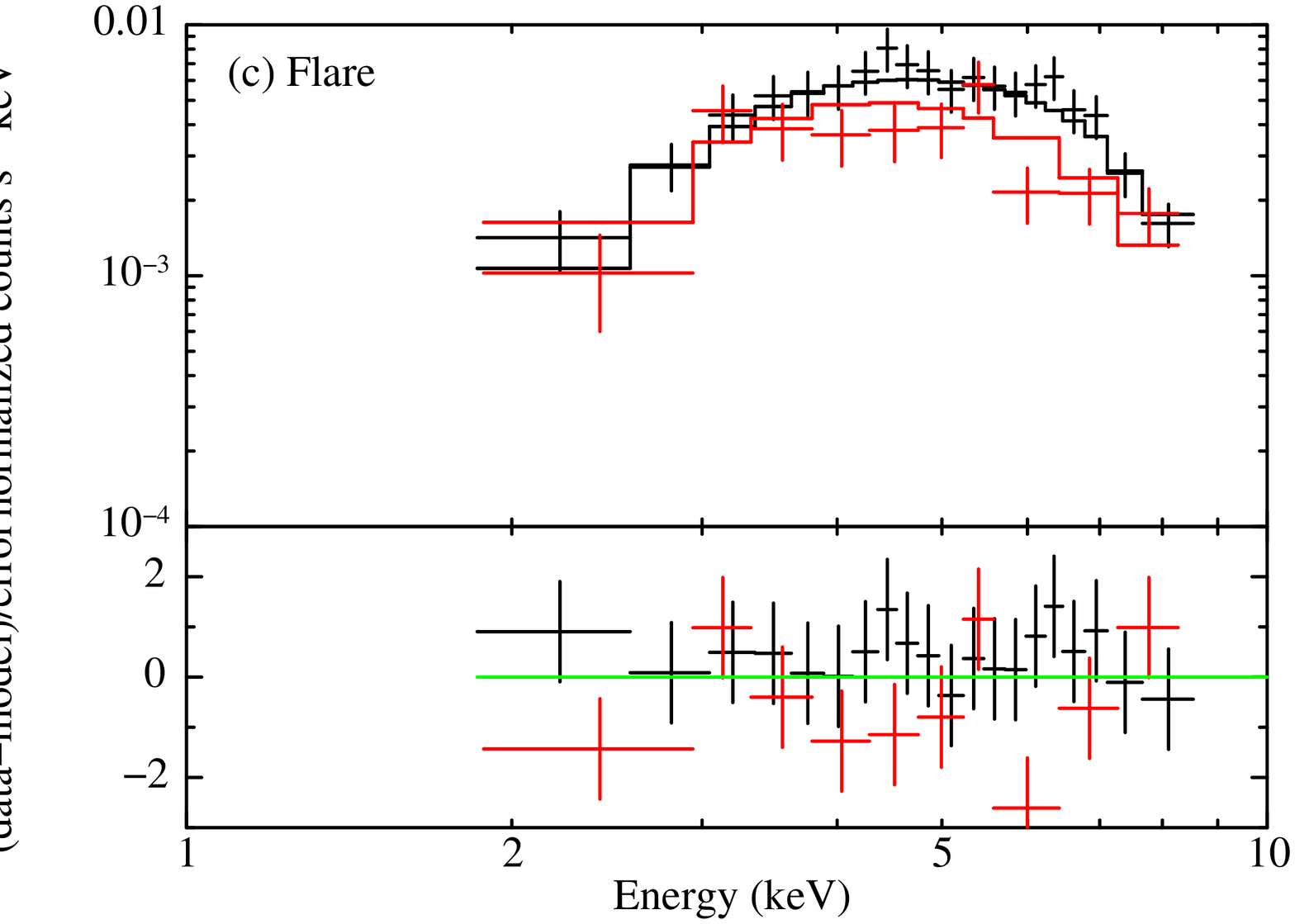}
\caption{Background-subtracted {\it Suzaku} spectra of AX~J1714.1$-$3912 extracted from the total observation time, i.e., ``Average"  (Panel a), from the ``Quiescent" time period (Panel b), and from the ``Flare" phase (Panel c). Black and red crosses represent the data set from the front-illuminated and back-illuminated chips, respectively. Solid lines shows the corresponding best-fit models.}
\label{fig:SuzakuSpec}
\end{figure*}

 \begin{table}
\caption{Best-fit parameters for the {\it Suzaku} spectra$^{a}$.}
% title of Table
\label{tab:SuzakuSpec}      % is used to refer this table in the text
\centering                          % used for centering table
\begin{tabular}{c c c c}        % centered columns (4 columns)
\hline\hline                 % inserts double horizontal lines
 & Average & Quiescence & Flare \\    % table heading 
\hline                        % inserts single horizontal line
${N_{\rm H}}^{b}$ &  8.8 (6.8--11.3) & 7.1 (3.6--12.2) & 8.3 (6.0--11.0) \\
$\Gamma$ & 1.4 (1.1--1.8) & 1.4 (0.6--2.3) & 1.2 (0.7--1.6) \\
${F_{\rm 2-10~keV}}^{c}$ & 1.8 (1.6--2.1) & 0.9 (0.7--1.2) & 5.4 (4.8--6.2) \\
${F_{\rm 2-10~keV}^{\rm obs}}^{d}$ & 1.1 & 0.6 & 3.6 \\
$\chi^2$/d.o.f. & 26.1/27 & 12.1/15 & 24.6/27 \\
\hline                                   %inserts single line
\multicolumn{4}{p{20pc}}{$^{a}$:Errors indicate single parameter 90\% confidence regions.}\\
\multicolumn{4}{p{20pc}}{$^{b}$:Absorption column
in the unit of $10^{22}$~cm$^{-2}$.}\\
\multicolumn{4}{p{20pc}}{$^{c}$: Intrinsic flux in the unit of
$10^{-12}$~erg~cm$^{-2}$s$^{-1}$
in the 2--10~keV band.}\\
\multicolumn{4}{p{20pc}}{$^{d}$: Observed flux in the unit of
$10^{-12}$~erg~cm$^{-2}$s$^{-1}$
in the 2--10~keV band.}
\end{tabular}
\end{table}

Figure~\ref{fig:SuzakuSpec}(a) shows the background-subtracted spectra of AX~J1714.1$-$3912. One can see that the emission is very hard and deeply absorbed, thus we fitted the spectra with an absorbed power-law model. For the absorption model, we applied the {\sc phabs} model, which includes the cross sections of \citet{balcinska-church1992}
with solar abundances \citet{anders1989}. The fitting was accepted with a $\chi^2=26.1$ (with 27 d.o.f.). Best-fit models and parameters are shown in Figure~\ref{fig:SuzakuSpec} and Table~\ref{tab:SuzakuSpec}, respectively. The spectral fitting confirmed the very large absorption column and the small photon index of the source.

In order to examine the spectral variations during the flare, we performed a time resolved spectral analysis and divided the observation time in two parts, corresponding to the ``Quiescence'' and ``Flare'' phases, as shown in Figure~\ref{fig:SuzakuLC}.
Source and background regions were the same as those adopted to extract the light curve.
Figure~\ref{fig:SuzakuSpec}(b) and (c) show the spectra in the Quiescence and Flare phases, respectively.
Both spectra are deeply absorbed and hard, and we fitted these spectra with the same model adopted to fit the total spectrum.
The best-fit parameters are summarized in Table~\ref{tab:SuzakuSpec}. None of the parameters change significantly during the flares except for the flux.
We note that the spectrum in the Flare phase shows positive residuals around the Fe-K line band, which may be associated with neutral iron line emission, similar to that observed  sometimes in the spectra of X-ray binaries. We then tried to add to our best-fit model a narrow Gaussian  with fixed central energy of 6.4~keV. However, the fitting was not improved, and we found an upper limit for the Fe line equivalent width of 175~eV.

Finally, we searched for coherent pulsations, but found nothing. This may be partially due to the relatively poor statistics and to the lack of time resolution of {\it Suzaku} XIS (8~s) \citep{koyama2007}.

\subsection{{\it Chandra}}
\label{chandra}

\begin{figure}[!htb]
\centering
\includegraphics[width=\columnwidth]{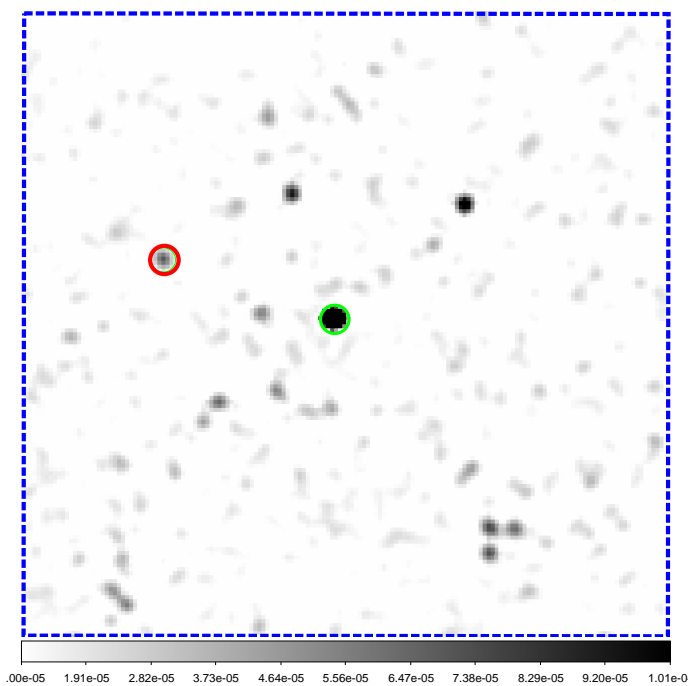}
\caption{{\it Chandra} ACIS-I count rate image of the AX~J1714.1$-$3912 region in the $0.5-8$ keV band . The grayscale is linear and the bin size is $1.968''$. The field of view corresponds to the blue dashed box indicated in Fig. \ref{fig:SuzakuImage}. Red and green circles indicate the position of the point-like sources CXOU J171343.9$-$391205 and CXOU J171236.7$-$391235, respectively.}
\label{fig:chandraImage}
\end{figure}

With respect to the previous $ASCA$ data, the {\it Suzaku} observation allowed us to obtain tighter constraints on the angular extension of the source (radius $<2'$, consistent with the {\it Suzaku} PSF), which was assumed to be much larger by \citet{uchiyama2002}. 
Figure~\ref{fig:chandraImage} shows the {\it Chandra} ACIS-I vignetting corrected count-rate image of the AX~J1714.1$-$3912 region in the $0.5-8$ keV band. The map clearly shows that the counterpart of the brightest source detected in the {\it Suzaku} observation is the relatively faint point-like source CXOU J171343.9$-$391205 (indicated by a red circle in Fig. \ref{fig:SuzakuImage} and Fig. \ref{fig:chandraImage}). The secondary source observed by {\it Suzaku} to the southwest can be associated with the point-like source CXOU J171236.7$-$391235 (green circle in Fig. \ref{fig:SuzakuImage} and Fig. \ref{fig:chandraImage}). 
As verified through dedicated simulations performed with the MIT/CXC MARX Chandra simulator, our {\it Chandra} observation would have been able to detect an extended stationary (i.e., with the same average flux as that observed with {\it Suzaku}) source with a Gaussian radial profile of surface brightness with $\sigma=2'$ with a signal-to-noise ratio $>10$ (much higher in case of a smaller source). The lack of extended emission in our {\it Chandra} data therefore definitely confirm that AX~J1714.1$-$3912 is indeed the point-like source CXOU J171343.9$-$391205.% which is characterized by highly variable X-ray emission.

We extracted the spectrum from CXOU J171343.9$-$391205 by collecting only 16 counts. We adopted the same spectral model as that used to describe the {\it Suzaku} Average spectrum, by freezing all the parameters but the normalization to the best-fit values reported in Table \ref{tab:SuzakuSpec}, and obtained an intrinsic X-ray flux $F_X=7\pm3\times10^{-14}$ erg~cm$^{-2}$~s$^{-1}$ in the $2-10$ keV energy band.

\begin{figure}[!htb]
\centering
\includegraphics[width=\columnwidth]{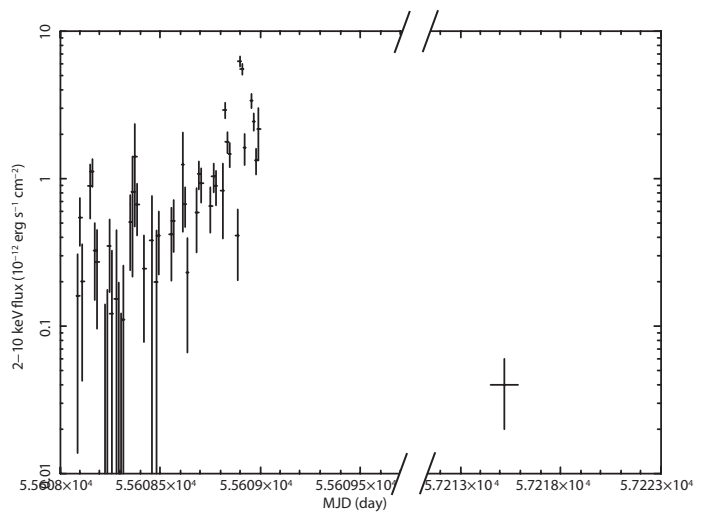}
\caption{Long-term light curve of AX~J1714.1$-$3912 in the 2--10~keV band obtained by combining {\it Suzaku} and {\it Chandra} observations.}
\label{fig:LC}
\end{figure}

\subsection{Long-term time variability}

We produced the long-term X-ray light curve by combining our {\it Suzaku} and {\it Chandra} observations. As in Sect. \ref{chandra}, we assumed that the spectral properties of AX~J1714.1$-$3912 (namely photon index and absorbing column density) are the same in the two observations to convert count rates to 2--10~keV observed flux.
Figure~\ref{fig:LC} shows the long-term light curve, which shows several flares in the {\it Suzaku} observation, whereas the X-ray flux drops down by 1--2 orders of magnitude  in the {\it Chandra} observation.

\section{Discussion and conclusions}
\label{sec:discussion}

We have discovered, for the first time, rapid time variability from the mysterious hard X-ray source, AX~J1714.1$-$3912. This is a strong clue that the source is point-like. Our {\it Chandra} dedicated follow-up observation has clearly confirmed the lack of diffuse X-ray emission in the region of AX~J1714.1$-$3912, which is instead associated with the point-like variable source CXOU J171343.9$-$391205. We can therefore exclude that the X-ray emission of AX~J1714.1$-$3912 that originates in high energy cosmic rays escaped from RX~J1713.7$-$3946 and diffused into a nearby molecular cloud.
In this section, we discuss the possible nature of AX~J1714.1$-$3912.

The time variability of the source has a timescale of a few thousand seconds, implying that the source is not extragalactic (e.g., an active galactic nuclei) but Galactic.
The absorption column that we measured from the X-ray spectra (see Table \ref{tab:SuzakuSpec}) is much higher than the Galactic absorption in the direction of the source \citep[(1.3--1.6)$\times 10^{22}$~cm$^{-2}$;][]{kalberla2005,dickey1990}, which therefore, has a high intrinsic absorption.
We searched for any counterparts in other wavelength with SIMBAD database and found nothing within 1~arcmin around both CXOU J171343.9$-$391205 and CXOU J171236.7$-$391235.

The intrinsic 2--10~keV luminosity is $L_{X}=1.4\times 10^{34}d_{8}^{2}$~erg~s$^{-1}$ during the {\it Suzaku} observation, where $d_8$ is the distance in units of 8~kpc.
The X-ray luminosity reached a maximum of $L^{max}_{X}\sim 5\times 10^{34}d_{8}^{2}$~erg~s$^{-1}$ during the Flare phase, and then became as small as $L^{Cha}_{X}\sim 3 \times 10^{32}d_{8}^{2}$~erg~s$^{-1}$ during the {\it Chandra} observation, performed about $4.5$~yr later. In the {\it ASCA} observations presented by \citet{uchiyama2002}, the intrinsic average luminosity was even higher, being $L^{ASCA}_{X}\sim 3\times 10^{35}d_{8}^{2}$~erg~s$^{-1}$.

Rapid time variability with a dynamic range of $> 10^2$ and with a timescale of a few thousand seconds suggests that AX~J1714.1$-$3912  may be a supergiant fast X-ray transient \citep[SFXT;][for review]{sidoli2011}. This association is consistent with its spectral properties, which are characterized by a hard and heavily absorbed X-ray emission.
The average luminosity of $10^{33-34}$~erg~s$^{-1}$ is also consistent with that of SFXTs. One of the SFXT, AX~J1841.0$-$0536, shows quasi-periodic flares \citep{bamba2001}.
%,which can be due to instabilities in the either magnetosphere \citep{lamb1977} or the accretion disk \citep{taam1984}.
The light curve is very similar to that of AX~J1714.1$-$3912, which may uphold the SFXT scenario.
Interestingly, another SFXT candidate (3XMM J185114.3-000002), with similar spectral properties, and similar short-term and long-term variability, was observed near another $\gamma-$ray emitting SNR, Kes 78, which is also interacting with a molecular cloud \citep{bth16,mbo17}.

The observed flux of AX~J1714.1$-$3912 is smaller than that of typical flares in SFXTs \citep{sidoli2011}. It is not clear whether AX~J1714.1$-$3910 is more distant compared with other SFXTs, or we just missed the peaks of large flares. On the other hand, the luminosity in the {\it Chandra} observation ($\sim 10^{32}$~erg~s$^{-1}$) is the smallest among the quiescent phases of SFXTs.
Our observation may provide a clue to the quiescent phase of SFXTs; SFXTs have very hard spectra up to $\sim$30~keV \citep{sidoli2011,walter2006}.
{\it Suzaku} Hard X-ray Detector (HXD) has a high sensitivity above 10~keV \citep{takahashi2007}
and has detected flares from several SFXTs \citep{kawabata2012,bth16}. In our case, however, the source was too faint to be detected with HXD.
Further observations in the hard X-ray band will be needed to understand the spectral shape of AX~J1714.1$-$3912.

It is believed that SFXTs are members of high mass X-ray binaries (HMXBs) with shorter outburst phases lasting a few days. Recently, \citet{pbp17} have suggested a possible way to distinguish SFXTs from classical supergiant HMXBs on the basis of the relationship between their absorbing column density and the Fe K$\alpha$ equivalent width. Our estimates of $N_H$ (Table \ref{tab:SuzakuSpec}) and our upper limit on the Fe equivalent width (Sect. \ref{Suzaku}) are consistent with an SFXT origin (see Fig. 1 in \citealt{pbp17}). However, these various hints are far from conclusive and better observational constraints are needed to get reliable distance estimates and to detect new flares (if any) beyond that presented here. In order to unambiguously conclude that AX~J1714.1$-$3912 is a new member of SFXTs, we need to detect an optical counterpart. Further follow-up observations will be encouraged to find this counterpart.

In conclusion, we analyzed {\it Suzaku} and {\it Chandra} observations of AX~J1714.1$-$3912 to ascertain the origin of its hard X-ray emission, which was originally associated with diffusion of extremely energetic particles from a SNR to a nearby molecular cloud. We verified that the source is point-like and highly variable both on short ($\sim 10^3$ s) and long (yr) timescales. We can therefore exclude any relationships between AX~J1714.1$-$3912 and the SNR RX~J1713.7$-$3946. The source is instead to be associated with CXOU J171343.9$-$391205 and both its spectral properties and time variability suggest that it is a new member of the SFXT class, although further observations are necessary.
% 
% \section{Conclusion}
% \label{sec:conclusion}

\section{Note after acceptance}
Following the advice of Dr. M. Petropoulou and Dr. G. Vasilopoulos, we inspected the IPAC-2MASS catalogue and noticed that the source 2MASS J17134391-3912055 matches the position of the flaring X-ray source CXOU J171343.9$-$391205. The presence of infrared emission further supports a possible association with an HMXB

\begin{acknowledgements}
We appreciate the comments and suggestions by the referee. The authors wish to thank Dr. S. Gabici for interesting discussions and comments.
The scientific results reported in this article are based to a significant degree on observations made by the {\it Chandra} and {\it Suzaku} X-ray Observatories. This research has made use of the SIMBAD database, operated at CDS, Strasbourg, France \citep{wenger2000}. This work is supported in part by Grant-in-Aid for Scientific Research of the Japanese Ministry of Education, Culture, Sports, Science and Technology (MEXT) of Japan, JP15K051017 (A.~B.). This paper was partially funded by the PRIN INAF 2014 grant ``Filling the gap between supernova explosions and their remnants through magnetohydrodynamic modeling and high performance computing" (M.M.).
\end{acknowledgements}

% 
% \begin{table}
% \caption{Observation Log}
% \label{tab:obslog}
% \centering
% \begin{tabular}{cccc}
% \hline\hline
%  & Obs start & Obs end & Exposure \\
%   & (YYYY/MM/DD HH:MM:SS) & (YYYY/MM/DD HH:MM:SS) & (ks) \\ \hline
% {\it Suzaku} & 2011/02/16 01:17:29 & 2011/02/16 23:10:11 & 32.6 \\
% {\it Chandra} \\
% \hline
% \end{tabular}
% \end{table}
% 


\begin{thebibliography}{}

\bibitem[{{Acero} {et~al.}(2009){Acero}, {Ballet}, {Decourchelle},
  {Lemoine-Goumard}, {Ortega}, {Giacani}, {Dubner}, \&
  {Cassam-Chena{\"i}}}]{abd09}
{Acero}, F., {Ballet}, J., {Decourchelle}, A., {et~al.} 2009, \aap, 505, 157
%
\bibitem[Anders \& Grevesse(1989)]{anders1989}
Anders, E., \& Grevesse, N. 1989, \gca, 53, 197
%
\bibitem[{{Arnaud}(1996)}]{arn96}
{Arnaud}, K.~A. 1996, in ASP Conf. Ser. 101: Astronomical Data Analysis
  Software and Systems V, 17
%
\bibitem[Balucinska-Church \& McCammon(1992)]{balcinska-church1992}
Balucinska-Church, M., \& McCammon, D.\ 1992, \apj, 400, 699 
%
\bibitem[Bamba et al.(2001)]{bamba2001}
Bamba, A., Yokogawa, J., Ueno, M., Koyama, K., \& Yamauchi, S.\ 2001, \pasj, 53, 1179 
%
%\bibitem[Bamba et al.(2016)]{bamba2016}
%Bamba, A., Terada, Y., Hewitt, J., et al.\ 2016, \apj, 818, 63 
%
\bibitem[{{Bamba} {et~al.}(2016){Bamba}, {Terada}, {Hewitt}, {Petre},
  {Angelini}, {Safi-Harb}, {Zhou}, {Bocchino}, \& {Sawada}}]{bth16}
{Bamba}, A., {Terada}, Y., {Hewitt}, J., {et~al.} 2016, \apj, 818, 63
%
\bibitem[{{Berezhko} \& {V{\"o}lk}(2007)}]{bv07}
{Berezhko}, E.~G. \& {V{\"o}lk}, H.~J. 2007, \apjl, 661, L175
%
\bibitem[{{Blandford} \& {Eichler}(1987)}]{be87}
{Blandford}, R. \& {Eichler}, D. 1987, \physrep, 154, 1
%
\bibitem[Butt et al.(2001)]{butt2001}
Butt, Y.~M., Torres, D.~F., Combi, J.~A., Dame, T., \& Romero, G.~E.\ 2001, \apjl, 562, L167 
\bibitem[Dickey \& Lockman(1990)]{dickey1990}
Dickey, J.~M., \& Lockman, F.~J.\ 1990, \araa, 28, 215 
%
\bibitem[{{Bykov} {et~al.}(2000){Bykov}, {Chevalier}, {Ellison}, \&
  {Uvarov}}]{bce00}
{Bykov}, A.~M., {Chevalier}, R.~A., {Ellison}, D.~C., \& {Uvarov}, Y.~A. 2000,
  \apj, 538, 203
%
\bibitem[{{Cassam-Chena{\"i}} {et~al.}(2004){Cassam-Chena{\"i}},
  {Decourchelle}, {Ballet}, {Hwang}, {Hughes}, {Petre}, \& {et al.}}]{cdb04}
{Cassam-Chena{\"i}}, G., {Decourchelle}, A., {Ballet}, J., {et~al.} 2004, \aap,
  414, 545
%
\bibitem[{{Fukui} {et~al.}(2012){Fukui}, {Sano}, {Sato}, {Torii}, {Horachi},
  {Hayakawa}, {McClure-Griffiths}, {Rowell}, {Inoue}, {Inutsuka}, {Kawamura},
  {Yamamoto}, {Okuda}, {Mizuno}, {Onishi}, {Mizuno}, \& {Ogawa}}]{fss12}
{Fukui}, Y., {Sano}, H., {Sato}, J., {et~al.} 2012, \apj, 746, 82
%
\bibitem[{{Gabici} {et~al.}(2009){Gabici}, {Aharonian}, \& {Casanova}}]{gac09}
{Gabici}, S., {Aharonian}, F.~A., \& {Casanova}, S. 2009, \mnras, 396, 1629
%
\bibitem[Kalberla et al.(2005)]{kalberla2005}
Kalberla, P.~M.~W., Burton, W.~B., Hartmann, D., et al.\ 2005, \aap, 440, 775 
%
\bibitem[Kawabata Nobukawa et al.(2012)]{kawabata2012}
Kawabata Nobukawa, K., Nobukawa, M., Tsuru, T.~G., \& Koyama, K.\ 2012, \pasj, 64, 99 
%
\bibitem[Koyama et al.(2007)]{koyama2007}
Koyama, K., Tsunemi, H., Dotani, T., et al.\ 2007, \pasj, 59, 23 
%
% \bibitem[Lamb et al.(1977)]{lamb1977}
% Lamb, F.~K., Fabian, A.~C., Pringle, J.~E., \& Lamb, D.~Q.\ 1977, \apj, 217, 197
%
\bibitem[Matsumoto et al.(2007)]{matsumoto2007}
Matsumoto, H., Ueno, M., Bamba, A., et al.\ 2007, \pasj, 59, 199 
%
\bibitem[Miceli et al.(2017)]{mbo17}
Miceli, M., Bamba, A., Orlando, S., Zhou, P., Safi-Harb, S., Chen, Y., \& Bocchino, F.\ 2017, \aap, 599, 45
%
\bibitem[Mitsuda et al.(2007)]{mitsuda2007}
Mitsuda, K., Bautz, M., Inoue, H., et al.\ 2007, \pasj, 59, 1 
%
\bibitem[{{Moriguchi} {et~al.}(2005){Moriguchi}, {Tamura}, {Tawara}, {Sasago},
  {Yamaoka}, {Onishi}, \& {Fukui}}]{mtt05}
{Moriguchi}, Y., {Tamura}, K., {Tawara}, Y., {et~al.} 2005, \apj, 631, 947
%
\bibitem[Pradhan et al.(2017)]{pbp17}
Pradhan, P., Bozzo, E., Paul, B.\ 2017, \aap, in press
%
\bibitem[{{Reynolds}(2008)}]{rey08}
{Reynolds}, S.~P. 2008, \araa, 46, 89
%
\bibitem[Sidoli(2011)]{sidoli2011}
Sidoli, L.\ 2011, Advances in Space Research, 48, 88 
%
% \bibitem[Taam \& Lin(1984)]{taam1984}
% Taam, R.~E., \& Lin, D.~N.~C.\ 1984, \apj, 287, 761 
%
\bibitem[Takahashi et al.(2007)]{takahashi2007}
Takahashi, T., Abe, K., Endo, M., et al.\ 2007, \pasj, 59, 35 
%
\bibitem[Uchiyama et al.(2002)]{uchiyama2002}
Uchiyama, Y., Takahashi, T., \& Aharonian, F.~A.\ 2002, \pasj, 54, L73 
%
\bibitem[{{Vink}(2012)}]{vin12}
{Vink}, J. 2012, \aapr, 20, 49
%
\bibitem[Walter et al.(2006)]{walter2006}
Walter, R., Zurita Heras, J., Bassani, L., et al.\ 2006, \aap, 453, 133 
%
\bibitem[Wenger et al.(2000)]{wenger2000}
Wenger, M., Ochsenbein, F., Egret, D., et al.\ 2000, \aaps, 143, 9 
\end{thebibliography}
\end{document}